\documentclass[aps,prb,twocolumn,groupedaddress]{revtex4}
\usepackage{graphicx}

\begin{document}

\title{Interface States and Anomalous Quantum Oscillations in Graphene Hybrid Structures}

\author{C. P. Puls}
\author{N. E. Staley}
\author{Y. Liu}
\email{liu@phys.psu.edu}

\affiliation{Department of Physics, 
The Pennsylvania State University, University Park, PA 
16802}

\date{\today}
\begin{abstract}

We report the first experimental study of continuous sheets of graphene possessing both one- and two-layer portions.  In a bulk hybrid structure featuring two large-area one- and two-layer graphene areas, our measurements in a magnetic field revealed the formation of two independent sets of Landau levels with drastically different energy scales and level spacings as expected as well as new features in the quantum oscillation originating from the interface and device configuration.  In edge hybrids featuring large two-layer graphene with two narrow one-layer graphene edges, we observed an anomalous suppression in the amplitude of the quantum oscillation of 2LG.  The suppression is interpreted as a consequence of the locking of the one- and two-layer graphene Fermi energies, the associated charge imbalance, and the emergence of chiral interface states, whose physical consequences are yet to be fully understood.

\end{abstract}

\pacs{}

\maketitle

\begin{center} \textbf{I. INTRODUCTION} \end{center}

Individually, monolayer graphene (1LG) and bilayer graphene (2LG) have been the foci of a great deal of experimental and theoretical work since the exfoliation of atomically thin graphitic sheets from bulk graphite was first demonstrated \cite{NovoFilms}.  1LG is prized for its unique two-dimensional band structure resulting from the honeycomb lattice of carbon atoms.   At the Fermi energy, the electron and hole bands meet at a single point and follow a linear dispersion $E(k) = \hbar v_F k$, where $\hbar$ is the Planck constant and $v_F \approx 1 \times 10^6$ $m/s$ is the Fermi velocity \cite{WallaceGraphite}.  2LG consists of two stacked layers of graphene shifted from each other by one atomic spacing.  Even though it is intrinsically a zero gap semiconductor just like 1LG, the electron and hole bands of 2LG follow a quadratic dispersion relation \cite{McCannBilayer}.  Asymmetrically charging the top and bottom layer of 2LG opens a gap at the Fermi level roughly equal to the potential difference \cite{McCannBilayerGap}.

Resulting from their unique band structures in two-dimensions, two novel unconventional integer quantum Hall effects (QHEs) have been discovered in 1LG and 2LG  \cite{NovoNature1, ZhangBerrys, CastroNetoDisGraph, NovoNature2}.  The QHE in 1LG was shown to exist even up to room temperatures \cite{NovoRoomT}, the origin of which is related to high mobility and large spacing between adjacent Landau levels, given by

\begin{equation} E_{n_1} = v_F \sqrt{2 e \hbar B n_1}, \end{equation} 

where $e$ is the electron charge, $B$ the magnetic field, and $n_1 =$ 0, 1, 2 ... the Landau level index \cite{NovoNature1, ZhangBerrys, CastroNetoDisGraph}.  For 2LG,

\begin{equation} E_{n_2} = \hbar \omega_c \sqrt{n_2(n_2-1)}, \end{equation}

where $\omega_c$ = $eB/m^*c$ is the cyclotron frequency, $m^*$ the effective mass of charge carriers, $c$ the speed of light, and $n_2 =$ 0, 1, 2 ... again the index \cite{NovoNature2}.  As a result, the spacing between Landau levels in 1LG can be many times larger than those of 2LG.  For example, at $B =$ 8 T, $E_{n_1} =$ (93 meV) $\sqrt{n_1}$ while $E_{n_2} =$ (12 meV) $\sqrt{n_2 (n_2 - 1)}$, assuming for the latter $m^* =$ 0.08$m_e$, a value consistent with experimental and theoretical values away from the charge neutral point \cite{McCannBilayerGap}.  Unlike a conventional two-dimensional electron gas (2DEG), both 1LG and 2LG possess a Landau level at zero energy.  In charge biased 2LG, however, the degenerate zero-energy Landau level (corresponding to $n_2 =$ 0 or 1) is split into two that are subsequently shifted to the top and the bottom of the bias-induced energy gap \cite{CastroBiasedBilayer, McCannBilayerGap}.  In a one- and two-layer graphene hybrid (see below), a continuous 2DEG featuring two distinct sets of Landau levels with very different energy scales is found.

\begin{figure}
\includegraphics[scale=0.23]{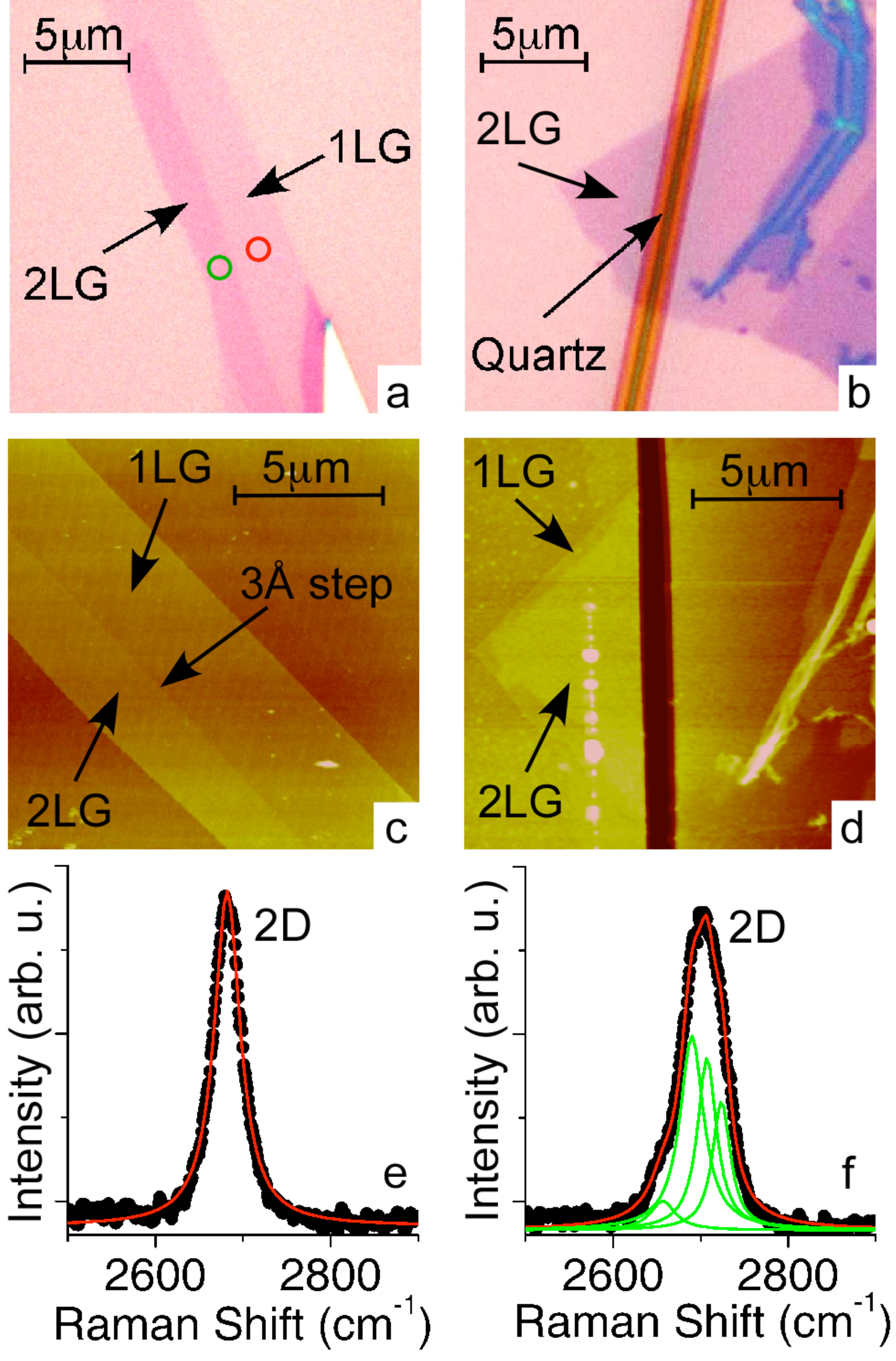}
\caption{ a) Optical image of a bulk hybrid graphene flake featuring large-area 1LG and 2LG portions.  Circles indicate approximate locations where Raman spectroscopy measurements were taken; b) Optical image of an edge hybrid with a quartz filament laid across it.  Narrow 1LG edges cannot be seen optically; c) AFM image of a bulk hybrid flake reveals a 3 \AA$ $ step at the interface; d) AFM image of an edge hybrid device with Au electrodes (taken after low-temperature measurements); Raman spectra of the 2D band of a hybrid flake confirm both a 1LG (e) and 2LG (f) portion.}
\end{figure}

Edge states in uniform 1LG and 2LG have been the focus of much theoretical study because of the important role they play in graphene transport \cite{ReviewGraphene}.  However, the step edge formed at the boundary of the neighboring regions of a continuous graphene sheet with different thicknesses has received little attention.  We are aware of only theoretical studies in which a semi-infinite 1LG was placed on top of an infinitely large 1LG.  Charge redistribution \cite{ArikawaEdgeStep} and enhanced local density of states at the step region \cite{CastroMultiEdge} were revealed. Electronic states at the interface were not examined.

Graphene also features high room temperature mobilities and unmatched mechanical strength, making it very interesting for potential practical applications which may require wafer-size, epitaxially grown graphene \cite{deHeerScience, deHeerEpi}.  Interface states and the associated charge distribution, which may produce conduction channels undesired for devices, are expected in epitaxial graphene whose thickness is not completely uniform on the micron scale.  Controlled studies of hybrid flakes will have important consequences for these grown films.

In this work, we study electronic transport parallel to the 1LG-2LG interface in graphene hybrid structures.  We observe a large charge imbalance across the interface between two systems with differently scaled electron energy levels in a magnetic field.

\begin{center} \textbf{II. EXPERIMENT} \end{center}

Our graphene flakes were mechanically exfoliated from bulk graphite and deposited onto a substrate of a heavily doped Si topped with 300 nm thick, thermally grown SiO$_2$.  Hybrid flakes, such as those shown in Fig. 1, were found among the graphite flakes deposited.  Thicknesses were determined by correlating optical characterization with Atomic Force Microscopy (AFM) and Raman spectroscopy measurements.  AFM measurements identified 3 \AA\ steps between the atomic layers of graphene in our hybrid geometry (Fig. 1c) and resolved regions of 1LG at the edges of 2LG even after Au leads ($\approx$ 200 \AA\ thick) were deposited (Fig. 1d).  However, mechanical exfoliation of graphene onto SiO$_2$ results in varying distances between the flake and substrate in different samples, possibly due to varying amounts of surface adsorbates.  Therefore, AFM cannot confidently measure the exact number of layers.  Raman spectroscopy is more conclusive in distinguishing between 1LG and 2LG; the 2D Raman peak of 1LG is fit to a single Lorentzian (Fig. 1e) and that of 2LG is fit to the sum of four \cite{FerrariRaman} (Fig. 1f).

Our devices were fabricated using an all-dry, lithography-free method, employing an ultrathin quartz filament as a shadow mask \cite{Staley, StaleyFluc} allowing for the fabrication of two-terminal devices.  For the junction to be well-defined, the quartz filament must be laid transverse to the hybrid interface so that we measure transport parallel to the edge.  The heavily doped Si was used as a backgate through which the number of charge carriers and thus the Fermi level of the graphene were tuned.  A negative change by 1 V in gate voltage, $V_g$, adds 7 $\times 10^{10}$ electrons per cm$^2$.   Electrical transport measurements were carried out in DC in a cryostat with a base temperature of 1.3 K and a magnetic fields up to 8 T.

We study two classes of hybrid devices: a bulk hybrid, which includes two large-area 1LG and 2LG portions and the interface between them (Fig. 2a), and edge hybrids, which are mostly 2LG but feature narrow 1LG edges (Figs. 2c and e).  In all samples, we probe graphene areas that are much wider (7-10 $\mu$m) than they are long ($<$ 1 $\mu$m).  The samples presented in this work featured field-effect mobilities between 1,000 and 6,000 cm$^2$/Vs (Table 1).  Mobility, $\mu$, was calculated by measuring the conductance $\sigma$ as a function of carrier density $n$ and employing the two-dimensional conductance formula, given by 

\begin{equation} \sigma_{2D} = \frac{ne^2\tau}{m^*} = en\mu, \end{equation}

where $\tau$ is the scattering time, leading to 

\begin{equation} \mu_{FE} = \frac{1}{e} \frac{\partial \sigma_{2D}}{\partial n} = \frac{1}{e \epsilon \epsilon_0} \frac{\partial \sigma_{2D}}{\partial V_g},  \end{equation}

where $\epsilon_0$ is the vacuum permittivity and $\epsilon$ is the dielectric constant of our gate dielectric.  In our case of SiO$_2$, $\epsilon$ = 3.9.  Therefore, mobilities were extracted from the slopes of $\sigma_{2D}$ vs. $V_g$ (Figs. 2b, d, and f), and were found to be different on either side of the charge neutral point in all samples, as observed previously in chemically doped graphene \cite{IBMAsymmetry}.  In the case of Edge hybrid 1, a transition from a higher mobility near the charge neutral point to a lower mobility far away was found (Fig. 2d).

\begin{figure}
\includegraphics[scale=0.23]{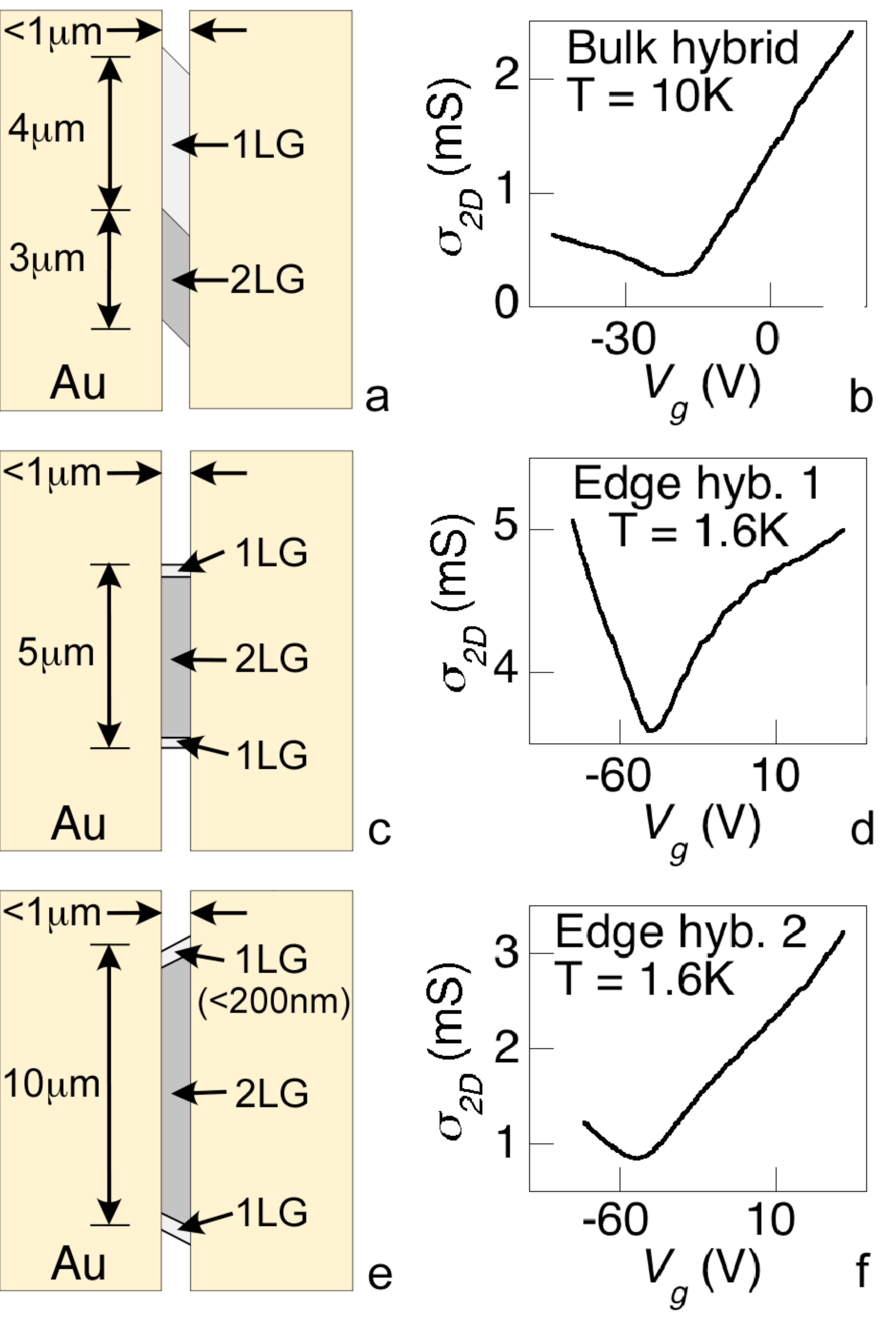}
\caption{a) Dimensions  and (b) conductivity ($\sigma_{2D}$) $vs.$ gate voltage ($V_g$) of the device prepared on the bulk hybrid in Fig. 1a; c) Dimensions  and (d) $\sigma_{2D}$ $vs.$ $V_g$ of the edge hybrid device referred to in FIg. 5a and 6c; e) Dimensions  and (f) $\sigma_{2D}$ $vs.$ $V_g$ of the device prepared on the edge hybrid in Fig. 1b.  A negative change in gate voltage corresponds to electron doping.  Extracted field-effect mobilities are listed in Table 1.}
\end{figure}

\begin{table}[b]
\caption{\label{tab:table1} Mobilities for electrons ($\mu_e$) and holes ($\mu_h$) for devices studied in this work}
\begin{ruledtabular}
\begin{tabular}{lcr}
Sample&$\mu_e$ (cm$^2$/Vs)&$\mu_h$ (cm$^2$/Vs)\\
\hline
Bulk hybrid& 6,000 & 1500\\
Edge hybrid 1& 1,000-2,500 & 4,500\\
Edge hybrid 2& 2,200 & 1,700\\
\end{tabular}
\end{ruledtabular}
\end{table}

\begin{center} \textbf{III. BULK HYBRID} \end{center}

Studying both bulk hybrids and edge hybrids give two perspectives on the effect of the interface on electronic transport.  The former is the most basic hybrid structure and the latter probes better the effect of the interface in a magnetic field, under which transport is dominated by channels at the edge of the sample.

Raman measurements of the bulk hybrid flake on SiO$_2$ in Fig. 1a before and after device preparation show a shift of the center of the 2D peak of about 4 cm$^{-1}$ indicating the sample was doped during device fabrication \cite{StampferRaman}, and necessarily top-doped.  The hybrid was electron-doped, most likely due to the diffusion of Au atoms onto the graphene during the deposition of electrodes.

\begin{figure}
\includegraphics[scale=0.23]{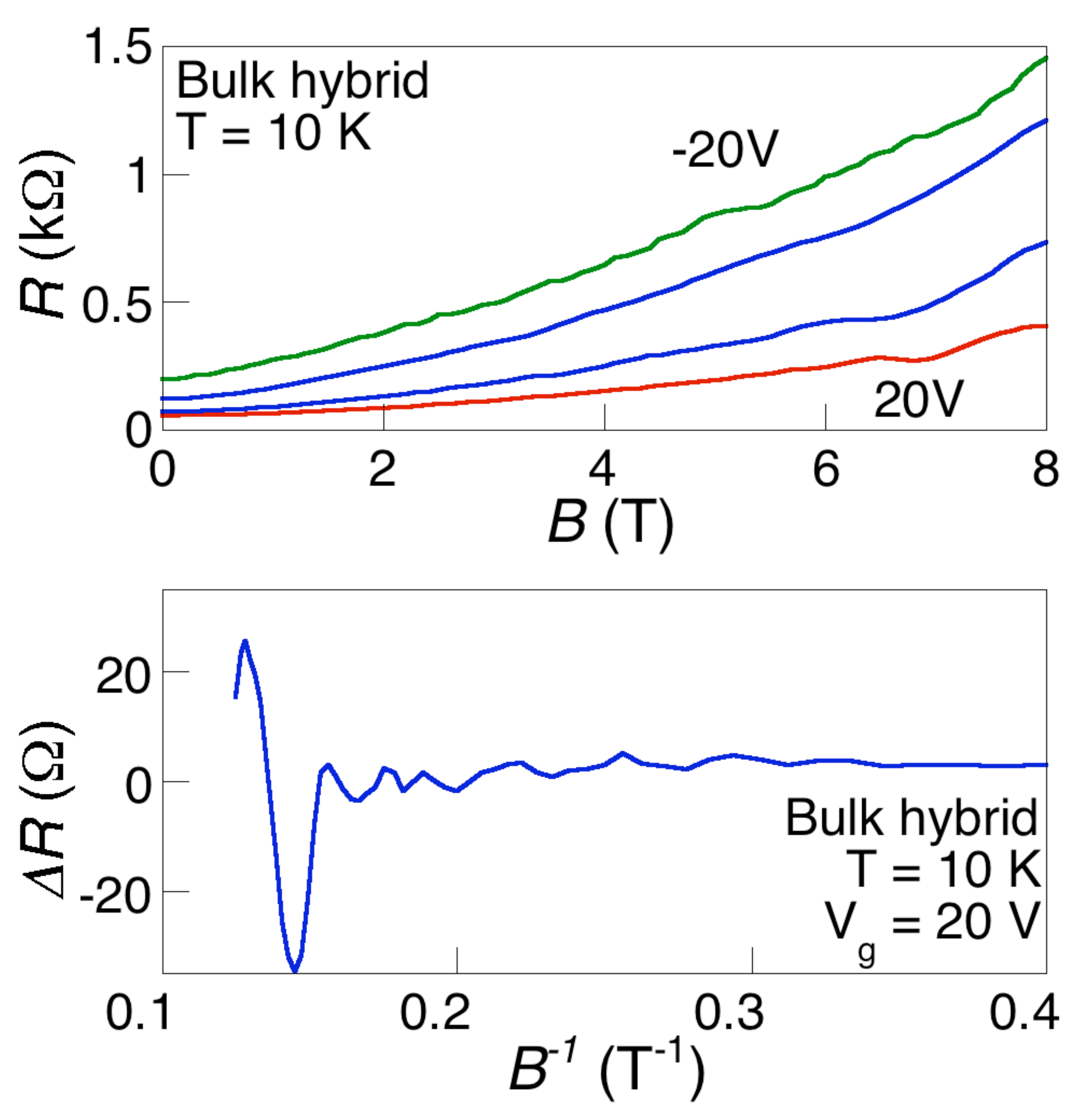}
\caption{a) Resistance ($R$) vs. magnetic field ($B$) at $V_g =$ -20, -12, 0, and 20 V for the bulk hybrid sample shown in Fig. 1a.  b) $R$ vs. $B^{-1}$ with a background removed reveals Shubnikov-de Haas Oscillations (SdHOs).  Inset: Fourier spectrum $\Delta R$ vs. $B^{-1}$.  We expect the presence of two sets of SdHOs, but their small amplitude and limited number makes evaluation of either difficult.}
\end{figure}

Large magnetoresistance ($\frac{\Delta R}{R} \approx 9-11$) was found at all gate voltages (Fig. 3) in a bulk hybrid, similar to what was found previously in pure 1LG \cite{FuhrerRvH}.  We expect quantum oscillations from both the 1LG and 2LG portions, but we could not confidently discern two separate periods from the Fourier transform of $\Delta R$ $vs.$ $B^{-1}$ (Fig. 3 inset) due to the limited number of oscillations observed.  In a constant magnetic field, the quantum oscillations were observed more clearly in $R$ $vs.$ $V_g$ (Fig. 4a).  To see the emergence of Landau levels in both 1LG and 2LG, the variation of the two-terminal conductance $\Delta \sigma$ obtained by subtracting a background is plotted as a function of $V_g$ in Fig. 4b for a magnetic field of $B =$ 6 T.  Given that the number of carriers any Landau level can accomodate is $\frac{fB}{\phi_0}$, where $f$ is the Landau level degeneracy ($f =$ 4 for both 1LG and 2LG) and $\phi_0 = \frac{2 \pi \hbar}{e}$ is the flux quantum, the expected period of $V_g$ is 8.3 V for $B =$ 6 T in both the 1LG and 2LG portions.  Furthermore, since Landau levels in 1LG (2LG) have an odd (even) integer filling factor \cite{ZhangBerrys, NovoNature1, NovoNature2}, levels should be filled alternately between 1LG and 2LG.  This would cause each set of oscillations to be shifted from one another by half a period.  Two such sets of were indeed found experimentally (Fig. 4b), marking the first observation of simultaneous quantum oscillations due to the presence of these two sets of unconventional Landau levels.  We note that in the bulk hybrid, the presence of alternating conductance peaks between 1LG and 2LG prevented well defined conductance plateaus from being observed.  

\begin{figure}
\includegraphics[scale=0.23]{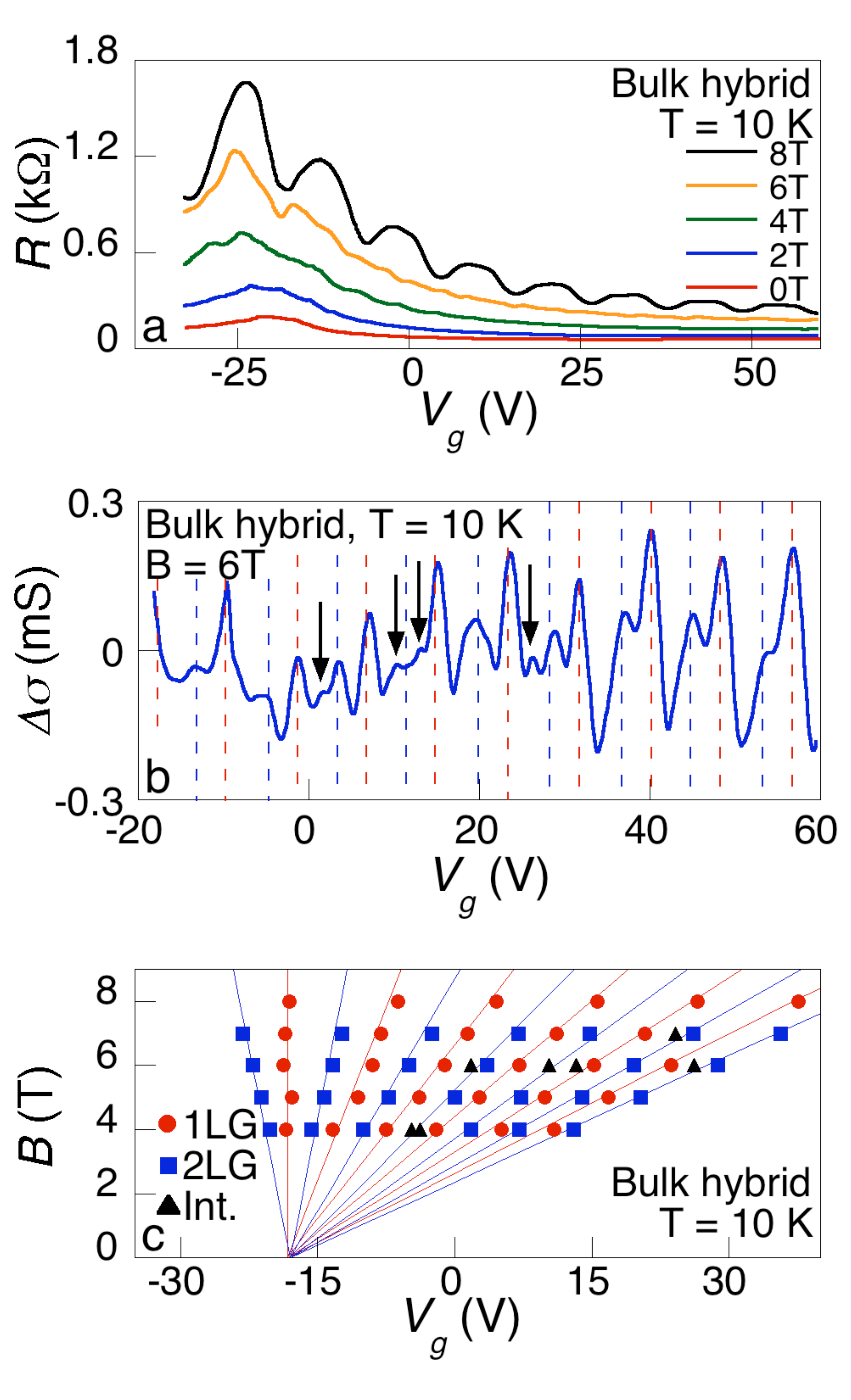}
\caption{a) $R$ vs. $V_g$ in various B-fields as indicated; b) Conductance variation $\Delta \sigma$ $vs.$ $V_g$ at $B =$ 6 T.  The red and blue dashed lines indicate where the Fermi level sweeps through Landau levels in 1LG and 2LG, respectively.  Black arrows point to weak, unexpected peaks possibly associated with the interface (see text); c) Positions of conductance peaks at $B$ = 4, 5, 6, 7, and 8 T, including the unexpected peaks (black triangles).  Higher oscillation amplitude of 1LG makes some 2LG peaks difficult to resolve and are unmarked on the diagram.  The red and blue lines show the theoretically expected positions of Landau levels in 1LG and 2LG, respectively, and are not fittings.  The charge neutral points of 1LG and 2LG are estimated to be at $V_g =$ -18.3 V and -18.0 V, respectively. }
\end{figure}

The geometry of samples (short and wide) results in conductance peaks as $V_g$ sweeps the Fermi level through peaks in the density of states at the positions of emergent Landau levels \cite{LevitovShape}.  While both the energy spectra of 1LG and unbiased 2LG have a Landau level at the charge neutral point whose position in $V_g$ would be field independent, only that of of 1LG has the adjacent levels spaced one integer filling factor away.  We found that only one set of oscillations that fit this requirement, and labelled it with red-dashed lines in Fig. 4b.  The other set (blue-dashed lines) necessarily belongs to 2LG and features a single integer filling factor separation between peaks on either side of the charge neutral point.  This implies that the double degeneracy of the zero-energy Landau level in 2LG was lifted, splitting it into two levels on either side of a charge bias induced energy gap.  Supposing the dopants in our device were deposited on the top layer during fabrication, an opened gap could be estimated to be as large as $\approx$20 meV based on the measured charge bias of 1.3 $\times 10^{12}$ electrons per cm$^2$ (corresponding to a charge neutral point found at $V_g$ = -18 for this device)\cite{McCannBilayerGap}.  While the existence of a true gap in 2LG devices is still uncertain experimentally, the splitting of the zero-energy Landau level to either side of the charge-neutral point has been observed previously \cite{CastroBiasedBilayer}.

The field-dependences of the peak positions are shown in Fig. 4c, along with their theoretical positions spreading out from the charge neutral points of the 1LG and 2LG portions.  The charge neutral point of 1LG was determined to be the average position of field-independent peak, and that of 2LG was the average position of the midpoints of the two peaks with lowest index.  The theoretical position in gate voltage of the center of the $i$th Landau level was determined by a filling factor equation,

\begin{equation} V_{LL,i} = 4 \frac{B}{\phi_0} \frac{\partial V_g}{\partial n} (i+m),\end{equation}

where $i =$ 0,1,2 ... is the Landau level index and $m$ = 0 for 1LG and 1 for 2LG (we remind the reader that we are tracking the positions of the middle of Landau levels rather than points of filled levels as is done traditionally).  The observed small offset from the expected half-period shift is due to a slight difference in doping levels of the 1LG and 2LG portions - the 1LG part is doped with about 1.28 $\times 10^{12}$ electrons per cm$^2$ while the 2LG is doped with 1.26 $\times 10^{12}$ electrons per cm$^2$.

Unexpected additional conductance peaks that did not follow any periodic trend were also found, as indicated by black arrows in Fig. 4b and black triangles in Fig. 4c.  While these conductance peaks were usually relatively weak, they can sometimes make resolving regular peaks impossible, such as the expected 2LG peak at around $V_g$ = 12 V in Fig. 5b.  These peaks may originate from the interface in the bulk hybrid, and will be revisited below.

\begin{figure}[b]
\includegraphics[scale=0.23]{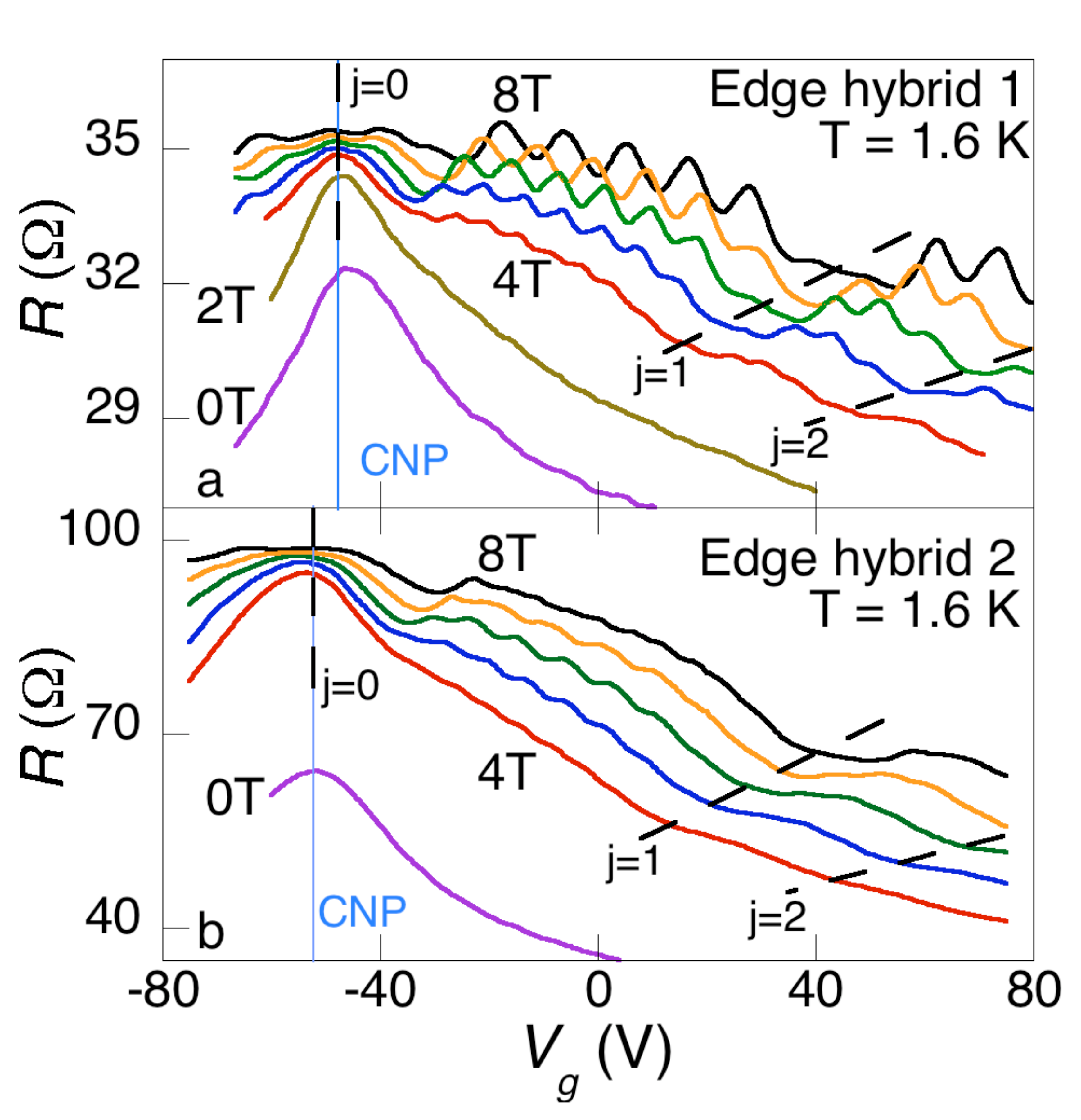}
\caption{a,b) Quantum oscillations in $R$ $vs.$ $V_g$ of two edge hybrid devices at $B =$ 0, 2, 4, 5, 6, 7, and 8 T ($B = $ 2 T data not taken for (b)).  Ranges of $V_g$-dependent suppression of 2LG resistance oscillation are indexed by $j$ (see text).   The 2LG portions of the edge hybrids feature resistance maxima at the charge neutral point (where doping has opened up a gap \cite{McCannBilayerGap}) and at integer filling factors of multiples of four, corresponding to DOS minima.  The resistance is unusually low due to a large width-length ratio of our devices.}
\end{figure}

\begin{center} \textbf{IV. EDGE HYBRIDS} \end{center}

Unlike a bulk hybrid, an edge hybrid features a large-area 2LG portion and narrow 1LG edges (Fig. 2c and e).  The presence of the hybrid interface the edge of the device (as opposed to the middle) results in very different behavior.

The flakes used, which were clearly 2LG based on our optical characterizations, were not imaged by AFM before the Au deposition because of the concern of possible damage of the flake by the AFM tip.  However, the AFM imaging after the low temperature measurements were carried out suggest that 1LG strips of a width less than 200 nm were present at the edges of the bulk 2LG.  

The edge hybrids were found to exhibit a regular resistance oscillation due to the filling of Landau levels of the large 2LG portions of the devices.  However, a dramatic suppression of the 2LG resistance oscillations was found over certain $V_g$ ranges (Fig. 5).  The center of these $V_g$ ranges over which the oscillations were suppressed (relative to the charge neutral point) varies roughly as $\sqrt{jB}$, where $j =$ 0, 1, 2, ... is the index of the oscillation suppression and $B$ is the magnetic field.  Interestingly, the energy of Landau levels in 1LG and 2LG are $E_{n_1} \sim \sqrt{n_1 B}$ and $E_{n_2} \sim \sqrt{n_2 (n_2 - 1)}$, where $n_{1,2} =$ 0, 1, 2, ... , respectively, which suggests that the suppression of the oscillations originates in the 1LG edges.

To show this connection, we first note that $E_{n_2} \sim n_2$ when $n_2$ is sufficiently large.  Given that each Landau level can hold an equal number of charge carriers, varying $V_g$, which increases the carrier density linearly, will change the Fermi energy, $E_F$, in 2LG linearly.  So long as $E_F$ of the 1LG is locked to that of 2LG in the edge hybrids, varying $V_g$ is operationally equivalent to varying the Fermi energy, thus making gate voltage essentially an energy ``ruler''.  The observed suppression of 2LG resistance oscillations could then be seen as a resistance oscillation of 1LG superimposed on those of the 2LG.

As pointed above, the energy scales for the Landau levels in the 1LG and 2LG are very different.  For example, the expected Landau level spectrum for $B =$ 5 T is shown in Fig. 6a for 1LG and charge biased 2LG with $m^*$ = 0.08$m_e$ (the breadths of the levels will be discussed below).  As $V_g$ is ramped up from the charge neutral point, the relatively large energy of the $n_1 =$ 1 Landau level in 1LG would encourage electrons in the 1LG edges to fill the lower-energy Landau levels in 2LG after the $n_1 =$ 0 level in 1LG is filled.  The charge imbalance between the 1LG and 2LG is maintained by a sufficiently large electric field across the interface.  The end result is a locking of the Fermi energy of the 1LG edges to that of the 2LG.

If the conductance variation in the edge hybrid can be made simply to be proportional the total DOS, we can simulate the total sample conductance variation as a function of  $V_g$ (Fig. 6b). Comparing the simulated and experimental conductance variation (obtained by subtracting a linear background from the sample conductance vs. $V_g$), we find striking similarity between the two as shown in Fig. 6c.  The anomalous quantum oscillations can be reproduced at each magnetic field near peaks corresponding to the particular Landau level indices in 2LG whose energies match levels in 1LG.

\begin{figure}[b]
\includegraphics[scale=0.17]{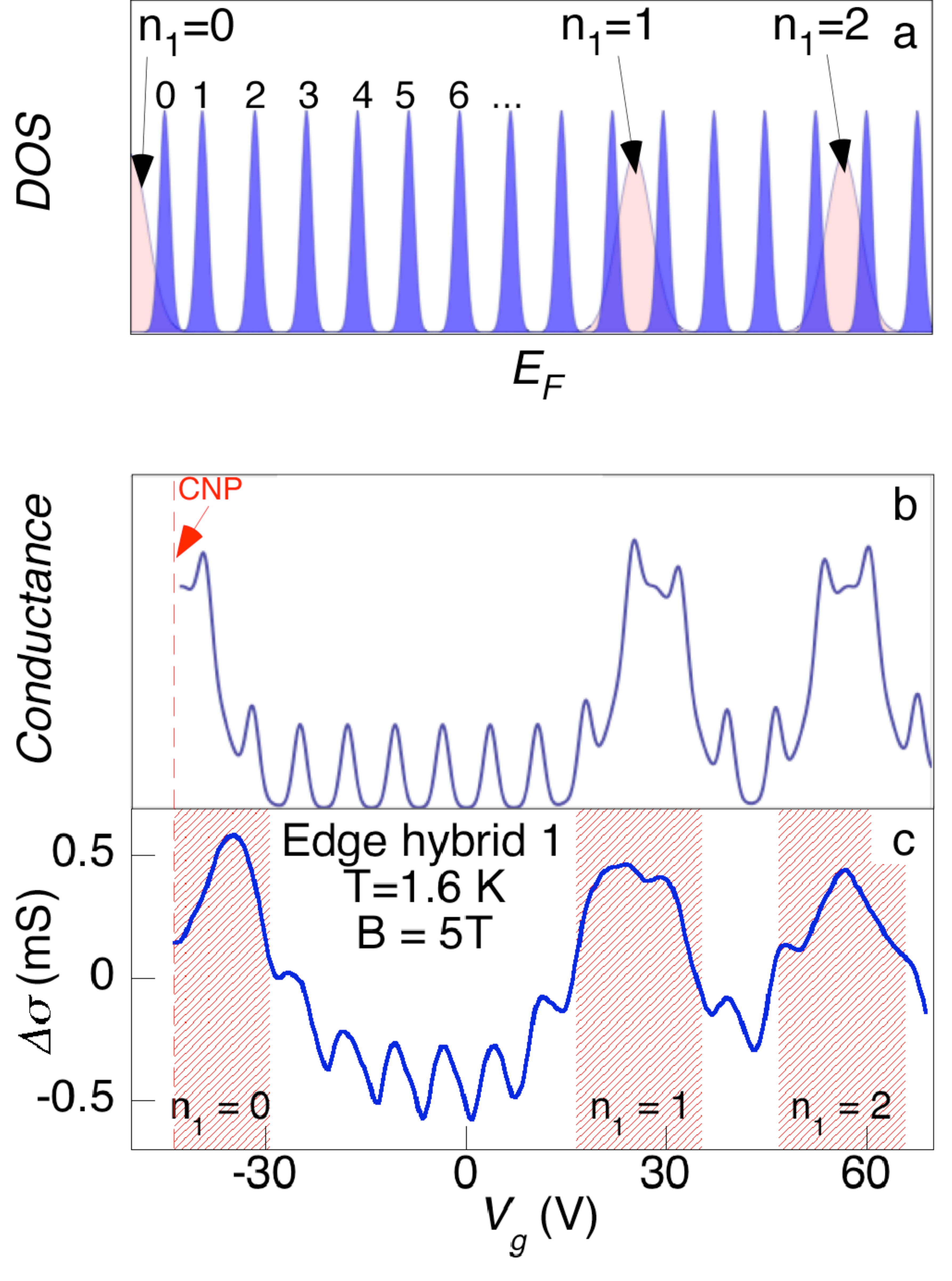}
\caption{a) Schematic of density of states $vs.$ Fermi energy ($E_F$) showing Landau levels in 1LG and biased 2LG (see text); b) Simulated conductance variation $vs.$ $V_g$ as determined by total density of states, assuming that $E_F$ of 1LG edges to that of 2LG; c) Experimental results of $\Delta \sigma$ $vs.$ $V_g$ for the edge hybrid device shown in the top of Fig. 4 for $B =$ 5 T. The shaded areas represent the variation in $E_F$ in 1LG edges due to factors described in text.}
\end{figure}

\begin{center} \textbf{V. DISCUSSION} \end{center}

The bulk hybrid featured, for the most part, behavior expected of independent 1LG and 2LG portions in parallel; the role of the interface was unclear.  This might be expected if the edges of the entire sample dominated the conductance channels in the presence of a perpendicular magnetic field, effectively ignoring the interface.  The edge hybrid, however, is a system where the hybrid interface is present at the edge of the device, allowing us to observe the effects of a charge imbalance across the interface.  We observed a locking of the Fermi energy of the 1LG edges to that 2LG, and the amount of charge involved was quite considerable.  At $B$=5 T, for example, the largest charge difference was $\approx$ 5 $\times 10^{12}$ electrons per cm$^2$ (the population of 10 Landau levels) when the Fermi energy is just below the $n_1 =$ 1 Landau level.  As to the width of the charge depletion region, it should be on the order of charge screening length, suggested by theoretical studies to be on the order of 10-100 nm \cite{ShklovskiiScreen, DasSarmaScreen}, consistent with that estimated from the size of charge puddles in 1LG \cite{YacobyPuddle}.  The regions of conductance enhancement spanned a wide range of $V_g$, implying that the degree of charge imbalance varied along the interface.  

The relative widths of the 1LG and 2LG contributions were considered in our simulation of conductance variation (Fig. 6c), and depend on several subtle factors.   Disorder is expected to be larger in the 1LG than in the 2LG of our edge hybrids because the edges of each graphene sheet are not necessarily smooth and tend to accumulate adsorbates.  Furthermore, inhomogeneity and the resulting variance of electric field across the interface will lead to variation in the gate voltage needed to place the Fermi energy at a Landau level.  We used a simple Gaussian peak to represent both of these complications in our simulation.

Other unusual features seen in resistance oscillations may also be attributed to interface states.  In edge hybrids, 2LG conductance peaks in regions of Landau level matching are missing or convoluted (Fig. 6b).  In bulk hybrids, small conductance peaks were found corresponding to neither 1LG nor 2LG (Fig. 4b).  These may be due to energy states and Landau level filling rates different from those in the bulk because of the charge redistribution.  Further studies of bulk hybrids measuring transport perpendicular to the interface may help resolve the effect of charge imbalances in these systems.  If an imbalance exists across the interface of charge neutral 1LG and 2LG in zero field due to, for example, a difference in work function, planar p-n junctions will then be expected.  1LG p-n junctions produced by local gates have been studied previously \cite{GoldhaberPN, MarcusPN, LevitovPN}.

The observation of the interface states and anomalous resistance oscillations in our hybrid structures demonstrates the rich physics of this unique graphene system where it is known that a 2DEG of Dirac fermions with a Berry phase of $\pi$ meets another unique 2DEG with a Berry phase of 2$\pi$ and a tunable energy gap.  In strong magnetic fields, the charge density in the interface region is expected to vary rapidly spatially, resulting in a strong lateral electric field, where the chiral edge states of 1LG and 2LG will meet. New interface states emerging from the coupling of these chiral states will have to be chiral as well, which should give rise to new interface phenomena.  In addition, the preferential charging of 2LG over 1LG may distribute unequally on the upper and the lower sheets of the 2LG, which may modify the 2LG chiral edge states in the context of a charge bias induced gap.  Finally, charge redistribution is to be expected at step edges in epitaxial graphene \cite{deHeerScience} where thickness may vary over a typical device length scale affecting the performance of the device.

\begin{center} \textbf{VI. CONCLUSION} \end{center}

We have measured electron transport in a magnetic field of two-point devices made on graphene hybrids - single, contiguous crystals consisting of both 1LG and 2LG portions and the interface between them.  Bulk hybrids feature quantum oscillations in magnetoconductance that correspond to the simultaneous filling of emergent Landau levels in 1LG and 2LG.  The interface appears to play only a subtle role in transport in a magnetic field, suggesting that the conductance in bulk hybrids is dominated by the edges of the whole sample rather than the edges of the individual 1LG and 2LG portions.  The effects of the 1LG-2LG interface in edge hybrids, however, is profoundly different.  In edge hybrid devices probing mostly 2LG flakes with narrow 1LG edges, anomalous quantum oscillations characterized by strongly enhanced conductivity suppress the regular 2LG oscillations at field-dependent intervals of gate voltage.  The anomalous oscillations are a result of the locking of the Fermi level in the 1LG edges to that of the bulk 2LG at the interface and the associated charge imbalance between two systems with very different electron energy scales in the presence of a magnetic field.  We have shown that the charge imbalance between 1LG and 2LG can be as much as 5 $\times 10^{12}$ electrons per cm$^2$ at the interface.

\begin{center} \textbf{ACKNOWLEDGMENTS} \end{center}

We would like to acknowledge the important contribution of A. Gupta, who helped with our Raman spectroscopy characterization of our graphene hybrids at the initial stage of this work.  We also  thank  J. K. Jain, K. von Klitzing, J. Zhu, P. Lammert, and J. Banavar for useful discussions.  This work was supported by DOE under DE-FG02-04ER46159 and DOD ARO under W911NF-07-1-0182.  

\bibliography{CPulsGrapheneHybridPRB}

\end{document}